# Carbon-Nanotube/β-Ga$_2$O$_3$ Heterojunction PIN Diodes

Hunter D. Ellis, Botong Li, Haoyu Xie, Jichao Fan, Imteaz Rahaman, Weilu Gao, and Kai Fu[a]

*Department of Electrical and Computer Engineering, The University of Utah, Salt Lake City, UT 84112, USA*

β-Ga$_2$O$_3$ is gaining attention as a promising semiconductor for next-generation high-power, high-efficiency, and high-temperature electronic devices, thanks to its exceptional material properties. However, challenges such as the lack of viable p-type doping have hindered its full potential, particularly in the development of ambipolar devices. This work introduces a novel heterojunction diode (HD) that combines p-type carbon nanotubes (CNTs) with i/n-type β-Ga$_2$O$_3$ to overcome these limitations. For the first time, a CNT/β-Ga$_2$O$_3$ hetero-p-n-junction diode is fabricated. Compared to a traditional Schottky barrier diode (SBD) with the same β-Ga$_2$O$_3$ epilayer, the CNT/β-Ga$_2$O$_3$ HD demonstrates significant improvements, including a higher rectifying ratio (1.2×10$^{11}$), a larger turn-on voltage (1.96 V), a drastically reduced leakage current at temperatures up to 300 °C, and a 26.7% increase in breakdown voltage. Notably, the CNT/β-Ga$_2$O$_3$ HD exhibits a low ideality factor of 1.02, signifying an ideal interface between the materials. These results underline the potential of CNT/β-Ga$_2$O$_3$ heterojunctions for electronic applications, offering a promising solution to current limitations in β-Ga$_2$O$_3$-based devices.

[a] Author to whom correspondence should be addressed: kai.fu@utah.edu

β-Ga$_2$O$_3$ has gained significant attention as an emerging semiconductor material for next-generation high-power and high-efficiency electronic devices due to its remarkable material properties. With an ultra-wide bandgap of 4.9 eV, a high critical breakdown electric field of 8 MV/cm, and an exceptional Baliga's figure of merit of 136 GW/cm$^{-2}$, β-Ga$_2$O$_3$ is particularly promising for high-power switching applications[1-4], enabling high breakdown voltages, low on-resistance, and low power losses[5]. These properties also enhance its resilience, enabling exceptional performance in high-temperature and radiation-rich environments[5, 6]. Recent advancements in growth techniques have significantly reduced production costs of β-Ga$_2$O$_3$ substrates, further its suitability in these applications[7-9].

Despite these advantages, the lack of viable p-type doping in β-Ga$_2$O$_3$ presents a significant challenge, as it hinders the development of ambipolar devices and limits the material's design versatility[10]. On the other hand, β-Ga$_2$O$_3$ based Schottky barrier diodes (SBD) have shown severely increased leakage current and reduced breakdown voltage at elevated temperatures due to the relatively low barrier height[5, 11, 12]. To address these limitations, heterojunction diodes (HDs) have been widely adopted by integrating a p-type material with n-type β-Ga$_2$O$_3$. Various p-type materials, including NiO, Cr$_2$O$_3$, and diamond, have been explored for forming the heterojunctions[13-19]. Some challenges of these materials have also been reported, such as hole concentration thermal stability of NiO[20], and low breakdown voltages of Cr$_2$O$_3$[15]. Researchers have been actively exploring more p-type materials for the HDs.

Two-dimensional materials offer another promising avenue due to their self-passivating properties[21]. Materials such as graphene, black phosphorus, and p-WSe$_2$ have been investigated, while graphene enables an exceptionally high critical electric field of 5.2 MV/cm[22-24]. Carbon nanotubes (CNTs) have also emerged as strong candidates for heterojunction applications, such as solar cells and photocatalysts, due to their outstanding properties, including high field-effect mobility, tunable bandgaps, and the capacity to sustain high current densities[25-33], making CNTs another potential p-type material for $Ga_2O_3$ heterojunctions.

In this work, we have studied the performance of a hetero-p-n-junction based on p-type CNT and i/n-type $Ga_2O_3$ for the first time. Key parameters, including ideality factor, barrier height, leakage current, and breakdown voltage, are systematically analyzed. Compared with Schottky diodes on the same epilayer, the CNT/β-$Ga_2O_3$ HD shows a larger turn-on voltage, larger rectifying ratio, reduced reverse leakage current up to 300 ℃, and increased breakdown voltage. The CNT/$Ga_2O_3$ diode also shows a low ideality factor of 1.02 compared to other HD diodes reported.

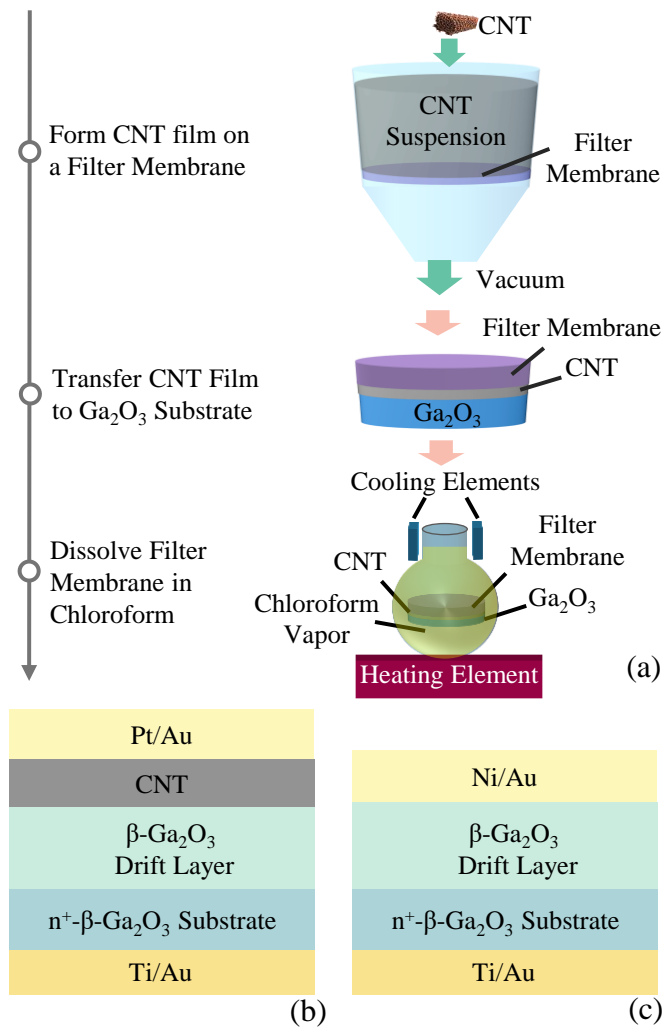

**FIG. 1.** (a) Schematics of the CNT deposition process flow. Structures of (b) the CNT/β-Ga$_2$O$_3$ HD and (c) the β-Ga$_2$O$_3$ SBD on the same epilayer.

The CNT thin film deposition was accomplished by first preparing a wafer-scale (2-inch diameter) randomly oriented CNT film using a vacuum filtration method[34]. Specifically, CNT powder was dispersed in a sodium deoxycholate (DOC) surfactant solution and subjected to tip sonication and ultracentrifugation to break down CNT bundles and remove impurities and undispersed bundles. The CNT suspension was then diluted while maintaining a DOC concentration higher than its critical micelle concentration. A fast vacuum filtration was performed to produce the film, which was left on the filtration system to dry completely. The dried film on the filter membrane was transferred onto the β-Ga$_2$O$_3$ using a wet transfer process[35]. A droplet of water was applied to the target substrate, and the CNT film was inverted, placing the CNT side in contact with the wet substrate while leaving the polycarbonate membrane above. Once the water evaporated, the sample was exposed to chloroform vapor inside a reflux condenser to dissolve and remove the polycarbonate membrane. The residual chloroform was removed by repeating the process with isopropyl alcohol, yielding a clean, randomly oriented CNT film. This CNT thin film deposition process is outlined in Fig. 1(a).

The β-Ga$_2$O$_3$ wafer was sourced from Novel Crystal Technology, Inc., and featured a 10 μm-thick drift layer with a doping concentration of $1 \times 10^{16}$ cm$^{-3}$. The β-Ga$_2$O$_3$ epilayer was grown on a (001) Sn doped n$^+$-β-Ga2O3 substrate. A Pt (50 nm)/Au (80 nm) alloy was sputtered to form ohmic contacts on the CNT film, and lift-off was used to remove the excess alloy. Pt was used instead of Ni for the CNT contact because Pt has less contact resistance with CNT. O$_2$ plasma was used to do a self-aligned mesa etch of the CNT. A back contact was formed by depositing Ti (50 nm)/Au (250 nm) on the Ga$_2$O$_3$ substrate surface. The CNT/β-Ga$_2$O$_3$ structure is shown in Fig. 1(b).

A Ga$_2$O$_3$ SBD was also fabricated as a reference for the CNT/β-Ga$_2$O$_3$ HD, and the structure is shown in Fig. 1(c). The same β-Ga$_2$O$_3$ wafer was used for the SBD fabrication as the HD. Ni(50 nm)/Au(100 nm) Schottky contacts were formed on the top surface, and lift-off was employed using the same photoresists as the HD to form the individual devices. The back contact was formed by depositing Ti (50 nm)/Au (250 nm). The forward *I-V*, along with the *C-V* measurements, were characterized using a 4200A-SCS parameter analyzer. The breakdown voltage was measured with a Keithley 2470 Source Meter.

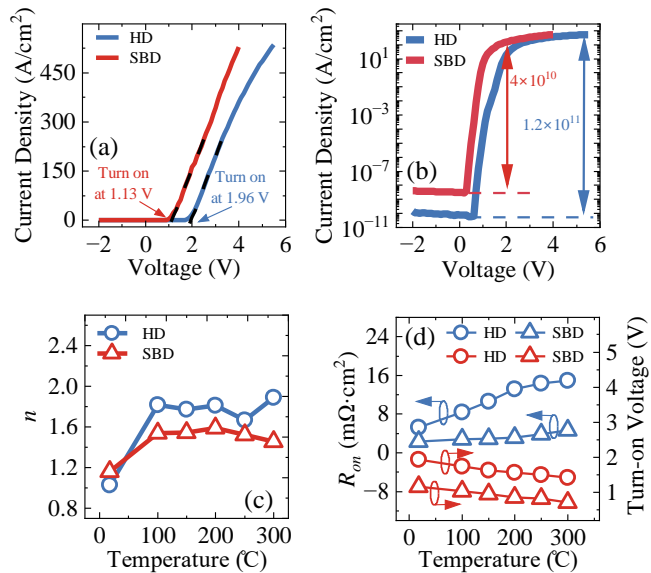

**FIG. 2.** Comparison of forward performance of the CNT/β-Ga$_2$O$_3$ HD and β-Ga$_2$O$_3$ SBD. (a) *I-V* curves in linear scale. (b) *I-V* curves in semi-log scale. (c) Ideality factor at various temperatures. (d) Differential on-resistance ($R_{on}$), and the turn-on voltages at different temperatures up to 300 ℃.

Figures 2(a) and (b) compare the forward *I-V* characteristics of the CNT/β-Ga$_2$O$_3$ HD and the SBD in both linear (Fig. 2(a)) and logarithmic (Fig. 2(b)) scales. The HD exhibits a higher turn-on voltage of 1.96 V compared to the SBD with a turn-on voltage of 1.13 V. This increase in turn-on voltage is indicative of an increased barrier height. Furthermore, the HD can achieve the same current density level as the SBD of 535 A/cm$^2$ at 5 V, demonstrating efficient carrier transportation across the heterojunction. The HD also shows significantly reduced reverse leakage current compared to the SBD, as evidenced in Fig. 2(b), further supporting the hypothesis of an increased barrier height or MIS structure formation. The similar on-current and reduced leakage current result in a high rectifying ratio of $1.2\times10^{11}$ for the HD, which is three times larger than the SBD, which has a rectifying ratio of $4\times10^{10}$.

Figure 2(c) presents the temperature dependence of the ideality factor ($n$) for both devices from 18 °C (room temperature) to 300 °C. At 18 °C, the HD exhibits a low ideality factor of 1.02, compared to the SBD with an ideality factor of 1.16. The ideality factor of nearly 1 for the HD demonstrates an ideal nature of the interface between the CNT and the β-$Ga_2O_3$. The ideality factor of the HD increases with temperature, consistent with the behavior observed in other β-$Ga_2O_3$ HDs, often attributed to effects such as interface states or tunneling[14]. Similarly, the SBD's ideality factor increases with temperature, potentially due to barrier inhomogeneities or interface states[36].

Figure 2(d) presents the differential on-resistance ($R_{on}$) and the turn-on voltage from 18 °C to 300 °C. The SBD exhibited a lower $R_{on}$ than that of the HD at all temperatures, a trend previously reported for NiO/β-$Ga_2O_3$ HDs[5,37,38]. Both devices show an increase in $R_{on}$ with temperature, due to the reduced carrier mobility in $Ga_2O_3$ due to phonon scattering[39]. The turn-on voltage decreases with temperature for both the HD and the SBD at a rate of 1.6 mV/K, which is due to the increased reverse saturation current density and reduced built-in potential resulting from the increased intrinsic carrier concentration at elevated temperatures. The dependence of the voltage on the reverse saturation current density is given in Eq. (1),

$$V = \frac{k_b T}{q} \times \exp(\frac{J}{J_0}+1) \qquad (1)$$

where $V$ is the applied voltage, $k_b$ is the Boltzmann constant, $T$ is the temperature, $J$ is the current density, and $J_0$ is the reverse saturation current[40].

The decrease in the bandgap of $Ga_2O_3$ with temperature could have also contributed to the decreased turn-on voltage[41, 42].

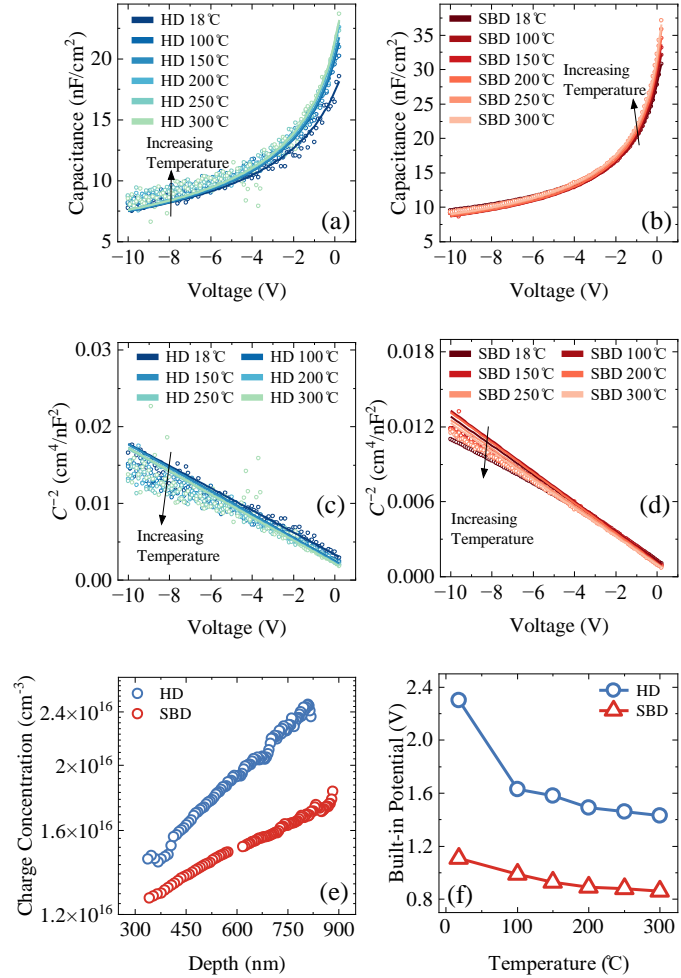

**FIG. 3.** *C-V* curves of (a) CNT/β-$Ga_2O_3$ HD and (b) the β-$Ga_2O_3$ SBD at different temperature. $C^{-2}$-*V* plots for (c) CNT/β-$Ga_2O_3$ HD and (d) the β-$Ga_2O_3$ SBD at different temperatures. (e) Derived doping concentration in the epilayers. (f) Derived built-in barrier height of the CNT/β-$Ga_2O_3$ HD and the β-$Ga_2O_3$ SBD at different temperatures.

Figure 3(a) and (b) display the *C-V* curves for the HD (Fig. 3(a)) and the SBD (Fig. 3(b)) from 18 °C to 300 °C. the HD shows smaller capacitance than the SBD due to the enhanced depletion depth in the HD at the same biases. Figure 3(c) and (d) present the $C^{-2}$-*V* characteristics of the HD

(Fig. 3(c)) and SBD (Fig. 3(d)) from 18 °C to 300 °C. The $C^{-2}$-$V$ curves reveal variations in the built-in potential ($V_{bi}$), with the HD exhibiting a $V_{bi}$ of 2.3 V and the SBD showing a $V_{bi}$ of 1.2 V at 18 °C, providing evidence of the increased barrier height in the HD. Figure 3(e) shows the doping concentration in the Ga$_2$O$_3$ epilayers at room temperature as a function of depth, derived from the $C$-$V$ data. The doping concentration was only determined for the low-doped β-Ga$_2$O$_3$ side due to the CNT's high intrinsic p-type doping concentration[43, 44]. While some differences are observed between the HD and SBD, they fall within the ±50% variation in doping concentration expected from the wafer manufacturer. The derived built-in potentials at different temperatures are presented in Fig. 3(f). The built-in potentials of both the HD and the SBD decrease with increasing temperature, which will cause the increase of leakage current at reverse biases[45].

Figure 4(a) compares the breakdown voltages of the HD and SBD. Without edge termination, the HD achieves a breakdown voltage of 912 V, significantly higher than the 723 V observed for the SBD. Additionally, the HD exhibits a significantly low leakage current, as illustrated in Fig. 4(a). The reverse $I$-$V$ curves from 0–200 V across the temperature range of 18 °C to 300 °C are presented in Fig. 4(b), and they highlight the HD's exceptional thermal stability, with leakage current at 300 °C smaller than the SBD at 18 °C. These results demonstrate the enhanced breakdown voltage, reduced leakage current, and improved thermal robustness of the CNT/β-Ga$_2$O$_3$ HD compared to the SBD.

The leakage mechanism of the SBD was explored, and it was found to be a combination of Poole Frenkel emission (PFE) and variable range hopping (VRH). The measured data compared to the simulated PFE and VRH current is shown in Fig. 4(c). The PFE leakage current was modeled using Eq. (2),

$$J_{PFE} = J_0 \exp\left(\frac{\beta E^{0.5}}{k_b T}\right) \quad (2)$$

where $J_{PFE}$ is the leakage current due to PFE, $\beta$ is the lowering of the barrier, and $E$ is the electric field at the interface[46].

The VRH leakage current was modeled using Eq. (3),

$$J_{VRH} = J_0 \exp\left(C \frac{qE}{2k_bT}\left(\frac{T}{T_0}\right)^{1/4}\right) \qquad (3)$$

where $J_{VRH}$ is the leakage due to VRH, $C$ is a constant, $q$ is the charge of an electron, and $T_0$ is the characteristic temperature[47].

The leakage mechanism of the HD was investigated by analyzing the dependence of the electric field on the leakage current based on the mechanisms listed in Table I. Figure 4(d) shows that the leakage current follows a relationship of $\log(I)/\log(E) \geq 2$, where $I$ is the leakage current and $E$ is the electric field. Based on Table I, the primary leakage mechanism of the HD is the space charge limited current (SCLC)[48]. Other possible mechanisms including PFE[46], and surface leakage[46], can also contribute to the leakage current. A comparison of key electrical parameters of the HD in this work and other reported $Ga_2O_3$ HDs are shown in Table II. The HD diode in this work achieves a very high rectifying ratio, the lowest ideality factor, a moderate $R_{on}$, and a decent turn-on voltage compared to the reported HDs with other p-type materials.

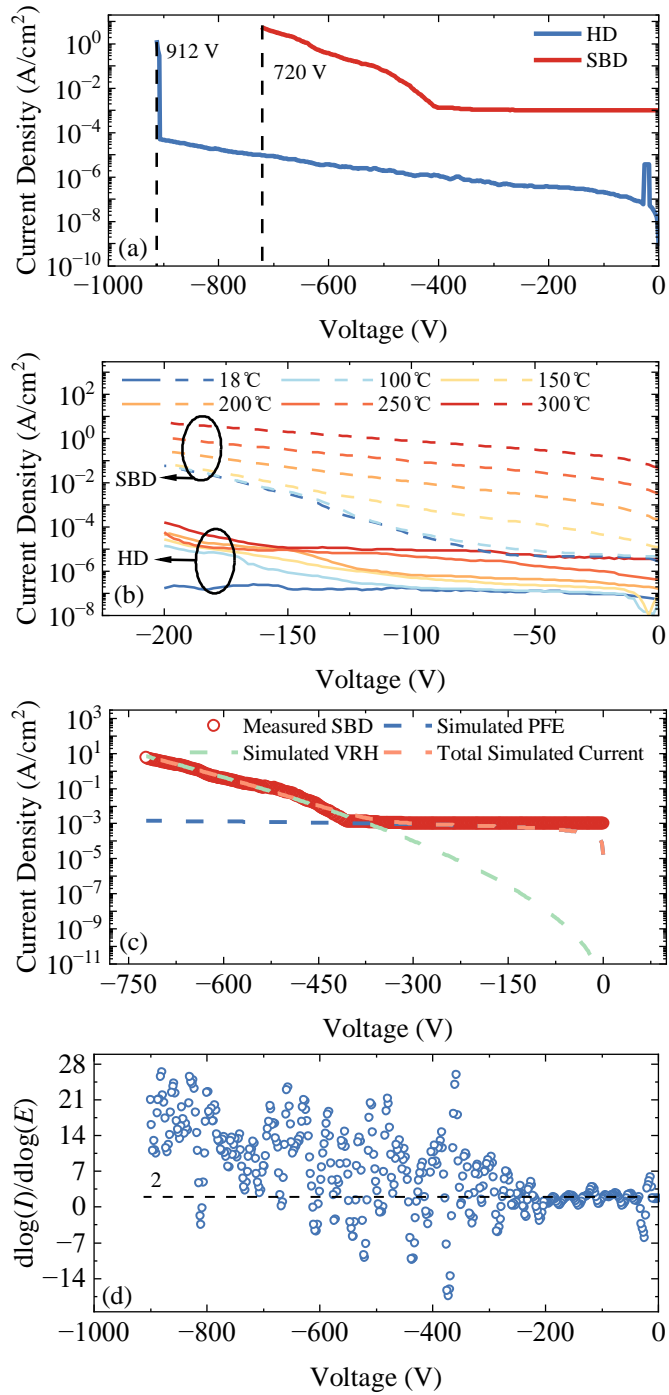

**FIG. 4.** (a) Reverse *I-V* curves and breakdown voltages of the CNT/β-Ga$_2$O$_3$ HD and the β-Ga$_2$O$_3$ SBD at room temperature. (b) Reverse *I-V* curves at high temperatures up to 300 °C. (c) Comparison of the measured SBD leakage current, simulated leakage current due to PFE, and

VRH mechanisms. (d) dlog(*I*)/dlog(*E*) vs voltage derived from the reverse *I-V* curve of the CNT/β-Ga$_2$O$_3$ HD.

Most research on CNTs focuses on their in-plane properties, while the vertical (out-of-plane) transport characteristics explored in this work have been rarely investigated. Another potential role of the CNT here contributing to the reduced leakage current could be the CNTs acting as an insulating dielectric. In this scenario, the forward current would rely on the carrier tunneling through the CNTs. Given the CNT layer's thickness (30 nm) and a calculated relative permittivity of approximately two (as shown in Fig. 5(a)), derived from *C-V* measurements taken on the device shown in the inset of Fig. 5(a) across frequencies from 1 MHz to 10 kHz, tunneling would produce negligible forward current while the forward current density achieved by the HD is almost the same as the SBD. So, we can confirm the p-type role of the CNT here. Furthermore, the experimental findings align well with the band diagram of a heterojunction p-n diode, as illustrated in Fig. 5(b). The configuration explains the increased barrier height, higher turn-on voltage, reduced leakage current, and comparable current density to the SBD.

**Table I.** Common leakage mechanisms in diodes.

| Mechanism | Equation | Differential Slope | Reference |
|---|---|---|---|
| SCLC | $I = \dfrac{9\varepsilon\mu E^n}{8W^3}$ $(n \leq 2)$ | $\dfrac{d\log(I)}{d\log(E)} \propto n$ | 48 |
| Poole-Frankel | $I = I_0 \exp(\dfrac{\beta E^{0.5}}{k_b T})$ | $\dfrac{d\log(\ln(I))}{d\log(E)} \propto 0.5$ | 46 |
| Surface Leakage | $I \propto \dfrac{E}{q}$ | $\dfrac{d\log(I)}{d\log(E)} = 1$ | 46 |

$\varepsilon$: permittivity, $W$: depletion width, and $\mu$: mobility.

**Table II.** Comparison of reported Ga$_2$O$_3$ HDs with different p-type materials.

| p-Material on Ga2O3 | Rectifying Ratio | Turn-on Voltage (V) | BV (V) | $R_{on}$ (mΩ/cm$^2$) | Ideality Factor | Reference |
|---|---|---|---|---|---|---|
| **CNT** | **1.2×10$^{11}$** | **1.96** | **912** | **5.2** | **1.02** | **This Work** |
| NiO | N.A. | 2.4 | 1059 | 3.5 | 1.22 | 16 |
| NiO | $10^{10}$ | 2.2 | 1860 | 10.6 | 1.8 | 49 |
| NiO | $10^{8}$ | 1.65 | 1630 | 4.1 | 1.27 | 14 |
| $Cr_2O_3$ | $10^{4}$ | 1.08 | 390 | 5.34 | 1.08 | 15 |
| NiO | $2.9\times10^{12}$ | 2.2 | 8000 | 7.8 | … | 50 |
| $CuO_x$ | $10^{8}$ | … | 2780 | 6.46 | 1.7 | 51 |

In conclusion, CNT/β-$Ga_2O_3$ HDs have been fabricated, demonstrating superior performance compared to a reference β-$Ga_2O_3$ Schottky barrier diode (SBD). The CNT/β-$Ga_2O_3$ HDs achieve a rectifying ratio three times higher, a 26.7% increase in breakdown voltage, a reduction in reverse leakage current by over two orders of magnitude, and an ideality factor close to 1. The advantage persists at elevated temperatures, with the HD consistently demonstrating significantly lower leakage current up to 300 °C The HD also matched the SBD in forward current density, reaching 535 A/cm², confirming efficient carrier transportation across the heterojunction. These findings establish CNT as a promising p-type material for β-$Ga_2O_3$ heterojunction diodes, offering significant potential for high-power and high-temperature applications.

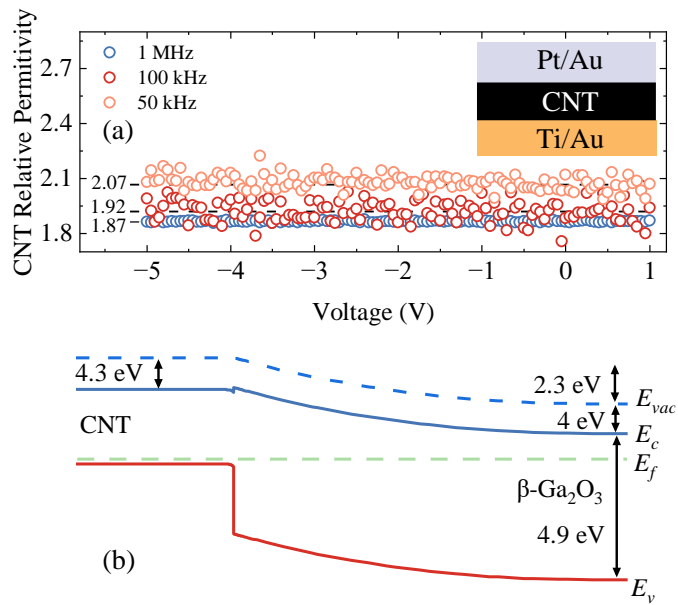

**FIG. 5.** (a) Relative permittivity of CNT from 10 kHz to 1 MHz, the inset shows the capacitor structure that was used to measure the *C-V*. (b) Band diagram for the CNT/β-$Ga_2O_3$ HD.

## AUTHOR DECLARATIONS

### Conflict of Interest

### Author Contributions

## DATA AVAILABILITY

The data that support the findings of this study are available from the corresponding authors upon reasonable request.

### Acknowledgment

The device fabrication was performed at the Nanofab at the University of Utah. This work is supported in part by the University of Utah start-up fund and PIVOT Energy Accelerator Grant U-7352FuEnergyAccelerator2023. X. H., J. F., and W. G. acknowledge the support from National Science Foundation through grants no. 2230727 and 2321366.